# Electron Paramagnetic Resonance of $V_N$-$V_{Ga}$ complex in BGaN

J. Kierdaszuk[1], E. B. Możdżynska[2], A. Drabińska[1], A. Wysmolek[1] , and

J.M.Baranowski[2]

[1]Faculty of Physics, University of Warsaw, Pasteura 5, 02-093 Warsaw, Poland

[2]Łukasiewicz Research Network-Institute of Microelectronics and Photonics, Al. Lotników 32/46, 02-668 Warsaw, Poland

**Abstract**

Metastable photoinduced Electron Paramagnetic Resonance (EPR) signal at low temperatures is reported in GaN alloyed with boron ($B_xGa_{1-x}N$) epitaxial layers grown at temperatures ranging from 840 °C to 1090 °C. An isotropic EPR line with g = 2.004 is observed with intensity depending on the growth temperature for all samples with boron content between 0.73% and 2.51%. Temperature dependence of EPR intensities is compared with the results of High-Resolution Photoinduced Transient Spectroscopy (HRPITS). This allows to link particular traps with EPR signal. The activation energies of these traps are consistent with the theoretical position of the $V_N$-$V_{Ga}$ complex. Thermal annihilation of the EPR signal with 30meV activation energy corresponds to shallow donor ionization. The model explaining light-induced EPR signal involving redistribution of electrons between deep and shallow donors mediated by photoionization to the conduction band is proposed.

## 1. Introduction

Electron paramagnetic resonance (EPR) proved to be a powerful technique for investigating and identifying defects in solids. The first work concerning EPR measurements in as-grown n-type GaN identified shallow donors.[1] The other example was the EPR studies of the radiation-induced defects in n-type GaN irradiated by 2 MeV electrons.[2] In this case, four defects were detected by EPR.[2,3] The well-studied defect was assigned to the oxygen gallium vacancy pair $(V_{Ga}O_N)^-$. There are also works that combine Optically Detected Electron Paramagnetic Resonance (ODEPR) with EPR or studies concerning magnetic dopants.[4–6]

To the best of our knowledge, no reports on EPR studies of BGaN are available. In this work, we present the results of EPR spectra in $B_xGa_{1-x}N$. The epitaxial layers measured in this work were the same as those used in previous investigations with a boron content ranging from 0.73% to 2.51%.[7,8]

Ternary boron gallium nitride (BGaN) has not been up to now used as a material which may expand bandgap engineering possibilities for future devices composed of III-N. The main challenge of boron-containing gallium nitride epitaxial growth is related to a low solubility of B in the GaN lattice. The upper solubility limit of B in the GaN lattice was found to be only a few percent.[7] It was shown that the BGaN epitaxial layers grown by MOCVD within the temperature range between 840 and 1090 °C show a decrease of boron incorporation into gallium sites ($B_{Ga}$) with increasing growth temperature.[7] On the other hand, the total amount of B determined by SIMS in these epitaxial layers was found to be constant for all growth temperatures.[7] This indicates that increased growth temperature leads to the transfer of B atoms from Ga substitutional sites to interstitial positions ($B_{inter}$) with the simultaneous creation of a gallium vacancy ($V_{Ga}$).[7] The reaction forming interstitial B is connected to the formation of Ga vacancies ($V_{Ga}$) and may go along the line proposed in recent theoretical work:[9]

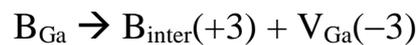
$$B_{Ga} \rightarrow B_{inter}(+3) + V_{Ga}(-3)$$

This reaction is more probable than postulated $B_{Ga} \rightarrow B_{inter}(+) + V_{Ga}(-)$ in our previous articles.[7,8] The activation energy for $B_{inter}$ formation is known to be relatively low, approximately 0.65 eV.[8] This activation energy indicates that the Fermi level is in a vicinity of the valence band.[9] Therefore, a very high concentration of $B_{inter}$ and $V_{Ga}$, of the order of $10^{20}$ cm$^{-3}$, is expected to be created at high growth temperatures in the investigated layers. The grown BGaN epilayers have a texture structure consisting of grains with a well-oriented c-axis. The $B_{inter}$ may easily diffuse out of the BGaN grains to the grain boundaries. Also, the $V_{Ga}$ will be mobile at growth temperatures.[10] They may diffuse out of BGaN grains to regions between the grains and form voids. Such voids have been observed in BGaN epilayers grown at high temperatures (Fig.1 in [7]). The presence of voids connected with Ga vacancies on the GaN:Si grains boundaries was also confirmed by positron annihilation.[11] However, some mobile $V_{Ga}$ defects may be stabilized within BGaN grains by forming complexes with positively charged defects or residual donors such as silicon or oxygen.[10]

The formation of these $V_{Ga}^-$ complexes may induce the generation of several traps in the vicinity of the top of the valence band. These traps have been reported by HRPITS and are responsible for a shift of luminescence into the red region.[8] In this report, we show that some of the $V_{Ga}$ complexes may be detected by EPR.

## 2. Experimental

The BGaN epitaxial layers were grown using 2-inch, single-side polished, (0001)-oriented $Al_2O_3$ substrates by MOCVD method as described in.[7,8] MOCVD growth was performed in AIX 200/4 RF-S low-pressure metalorganic vapor phase epitaxy reactor. The precursor gases used for BGaN growth were trimethylgallium (TMGa) and triethylboron (TEB). Ammonia ($NH_3$) and hydrogen ($H_2$) were used as carrier gases. The process of BGaN epilayers growth was carried out at four temperatures: 840 °C, 940 °C, 1040 °C and 1090 °C, and at the same pressure of 200 mbar. The boron content in obtained samples was 0.73%, 0.89%, 1.64 %, and 2.51%, respectively.

EPR measurements were performed using a Bruker ELEXSYS E580 spectrometer operating at a microwave frequency of 9.4 GHz (X-band) with a ER 4116DM resonance cavity. It was equipped with a liquid helium flow cryostat (Oxford ER 4112HV), allowing to cool the sample down to 5 K. During the measurements, the microwave power and the modulation amplitude were set to 0.47 mW and 0.1 mT, respectively. The optical window in the cavity enabled illumination of the sample by the xenon lamp with an absolute power of 80 watts. Sample heating by lamp illumination was estimated to be lower than a few Kelvin.

## 3. Results

The EPR spectra in all BGaN epilayers have been detected at low temperatures only under illumination. Figure 1 shows the evolution of the ESR spectrum for one of the samples, during the illumination by infiltered light from the Xe lamp. Initially, without illumination, no EPR signal was present. After a few minutes of illumination, the weak signal of the g-factor around 2.0040 became clearly visible. Although the signal-increasing rate is less noticeable over time, no saturation effect was observed even after 5 hours (Fig. 1). An experiment carried out using a 325 nm laser showed a similar effect.

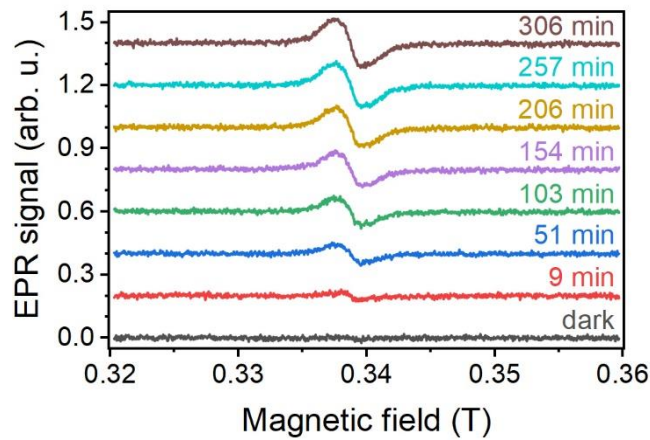

*Figure 1. Evolution of the EPR spectrum of the BGaN sample grown at 1090 °C as a function of the illumination time, measured at 5.4K.*

However, attempts using 440 nm and 532 nm lasers or bandpass filters removing the UV part of the Xe spectrum have not induced any EPR signal. This leads to the conclusion that interband (or close to interband) excitation is necessary to observe an increase in the EPR signal intensity. The low efficiency of this process is responsible for the slow excitation rate and lack of a saturation effect even after a long time of the experiment. The Xe lamp experiment was performed under the same conditions for four samples grown at different temperatures (Fig. 2). Each signal spectrum evolved similarly to that presented in Fig. 1, however, the rate at which signal intensity increases differs. No sign of saturation in any of investigated samples was observed.

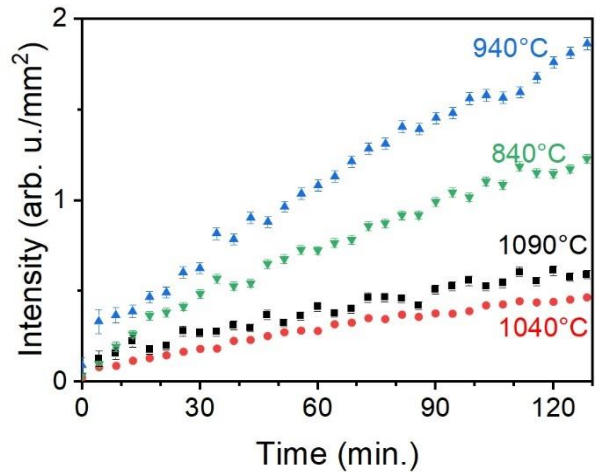

*Figure 2. Evolution of normalized (to the sample's surface area) EPR signal during illumination time for BGaN samples grown at different temperatures.*

The highest intensity is observed for the sample grown at 940 °C while the lowest is measured for the sample grown at 1040 °C. An additional experiment showed that the signal is stable after cessation of Xe lamp illumination, indicating the metastable character of the observed effect. For further studies, the experiment consisting of several annealing cycles was performed for samples grown at 840 °C and 1090 °C. Its motivation was to estimate the activation energy ($E_A$) at which the signal will decrease. The procedure was as follows: the sample was cooled in darkness to 5.4 K, then it was illuminated for 2 hours and the ESR spectrum was measured. Next, the sample temperature was increased to a certain temperature ($T_{anealling}$) and cooled down again to 5.4 K. Therefore, contrary to typical measurements performed as a function of temperature, each spectrum presented in Fig. 3 was collected at the same temperature (5.4 K).

Temperatures on the *x*-axis indicate the annealing temperature in the subsequent cycles. For the analysis of obtained data, we used a simple Arrhenius model described by the formula:

$$A(T) = A_0 e^{\frac{E_A}{k_B T}}. \tag{1}$$

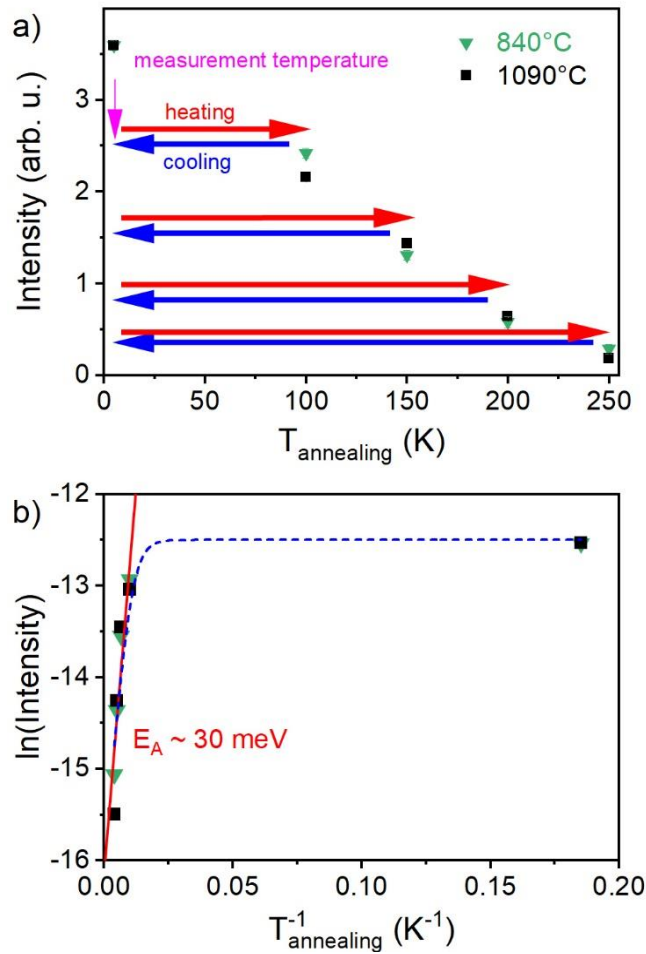

*Figure 3. Results of the annealing cycle experiment: a) the dependence of the ESR signal intensities (measured at 5.4 K) as a function of the annealing temperature (signals of both samples were normalized to 5.4K for better comparison) and b) the dependence of the natural logarithm of the EPR signal on the inverse temperature. The red arrows in figure a) indicate the way of sample annealing, while the blue arrows indicate the sample cooling in each cycle. The red line in figure b) indicates the fit of the activation model (1) to the experimental data, while the blue dashed line is the general trend.*

The results for samples 840 °C and 1090 °C are consistent and a steadily decrease of the signal intensity at annealing temperatures above 100 K was observed (Fig. 3a). This behavior was not observed for lower temperatures where the EPR signal was stable. Further analysis using equation (1) showed that the estimated activation energy of the signal-decreasing process is 30 meV (Fig. 3b).

## 4. Discussion

The origin of the observed EPR signal with an isotropic g = 2.004 in the investigated BGaN epitaxial layers cannot be directly identified. It is only possible to suggest the origin of the defect responsible for the observed EPR signal. The difference in EPR signal intensities between individual samples reflects differences of the concentration of defects present in the BGaN epilayers grown at different temperatures.

The recent publication of HRPITS measurements of the same BGaN samples allowed to find the energies and concentration of all traps in the layer and their dependence on growth temperature.[8] There are two traps T2 and T3 in which changes of concentration behave in a similar way as the EPR signal as shown in Fig.4. It strongly suggests that both Trap 2 and Trap3 are responsible for the presence of the EPR signal. Trap 2 and Trap 3 have energies of 0.64 eV and 0.78 eV above the valence band, respectively.[8] These energies are very close to the levels of two charge states of the nitrogen–gallium vacancy complex ($V_N$-$V_{Ga}$), which are 0.68 eV and 0.81 eV above the valence band maximum.[12] It has been shown that nitrogen-vacancy may exist in two charge states $V_N(+)$ and $V_N(3+)$.[9] Therefore, the complex $V_N(3+)$ with negatively charged $V_{Ga}$ will be relatively stable and may be responsible for the formation of Trap 2 and Trap 3 depending on the charge state of this complex. The complex $[V_N - V_{Ga}]^{2+}$ will have an even number of electrons and will not be paramagnetic. However, after photoionization, it will be changed to $[V_N - V_{Ga}]^{3+}$, which will have an unpaired electron, spin ½, and becomes EPR active.

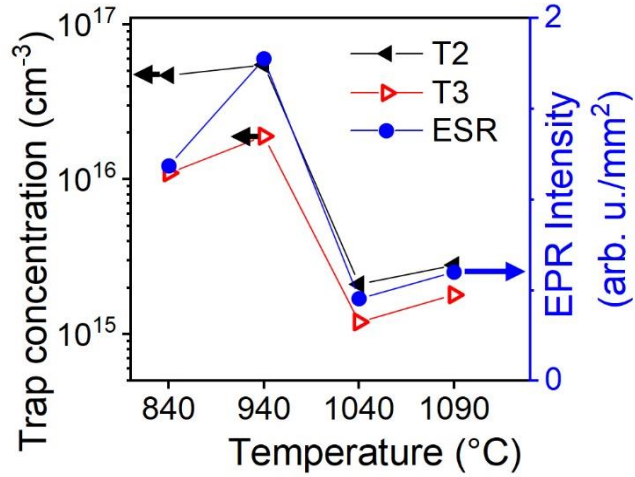

*Figure 4. The dependence of the concentration of T2 and T3 traps established in [4] compared with the normalized EPR signal intensities of BGaN epitaxial layers on growth temperature.*

The model of a creation of a paramagnetic center by light is shown in Fig. 5. Photoionization of the $[V_N-V_{Ga}]^{2+}$ complex will lead to an electron in the conduction band and the complex $[V_N-V_{Ga}]^{3+}$. This charge state of the complex will be in a metastable state at low temperatures. Electrons from the conduction band will be captured by residual donor centers such as silicon or oxygen, which for GaN are close to 30 meV.[11,13] The trapped electron by a donor center will blocate approach to the $[V_N-V_{Ga}]^{3+}$ complex. Recombination of the electron with the $[V_N-V_{Ga}]^{3+}$ complex requires thermal activation energy to release the trapped electron. Therefore, recovery of a stable nonparamagnetic charge state of the $[V_N – V_{Ga}]^{2+}$ complex will require an activation energy of 30 meV observed in the experiment.

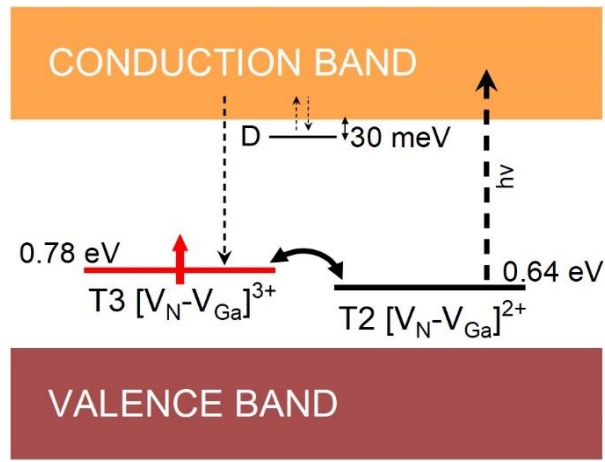

*Figure 5. Proposed model of illumination-assisted electron transition between nitrogen vacancy-gallium vacancy complexes in BGaN.*

## 5. Conclusions

Measurements of photoinduced metastable EPR signal with g = 2004 isotropic line in BGaN are presented. The intensity of the EPR signal in BGaN epitaxial layers grown at a temperature ranging from 840 °C to 1090 °C is compared with the results for trap energies and concentrations obtained in the HRPITS measurements for these samples. This comparison allows us to propose a model for the complex responsible for the photogeneration of the paramagnetic state. The model proposes photoionization of the non-paramagnetic $[V_N–V_{Ga}]^{2+}$ complex and generation of the paramagnetic $[V_N-V_{Ga}]^{3+}$ one. The identification of complexes is based on theoretically calculated energies[12], and experimental values of trap energies obtained in HRPITS measurements. The model explains the metastable character of the paramagnetic charge state of $[V_N-V_{Ga}]^{3+}$ and the mechanism of recovery of the $[V_N–V_{Ga}]^{2+}$. The mechanism of recovery is connected with an activation energy of 30 meV, which is related to the trapping of electrons by residual shallow donors.

**Acknowledgments**

Co-authors from the Łukasiewicz Research Network-Institute of Microelectronics and Photonics thank for the financial support from the Grant No. 4/Ł-IMIF/CŁ/2021 funded by The Łukasiewicz Centre. We also thank Prof. Roman Stępniewski for the valulable discussions.

**Data Availability**

The data that support the findings of this study are available from the corresponding author upon reasonable request.